# Phase Separation of Bismuth Ferrite into Magnetite under Voltage Stressing

X. J. Lou,[1*] C. X. Yang,[2] M. Zhang,[1] and J. F. Scott[1§]

[1]Centre for Ferroics, Department of Earth Sciences, University of Cambridge, Cambridge CB2 3EQ, U. K.

[2]Asic&System State Key Lab, Department of Microelectronics, Fudan University, P.R. China

**Abstract**

Micro-Raman studies show that under ~700 kV/cm of d.c. voltage stressing for a few seconds, thin-film bismuth ferrite $BiFeO_3$ phase separates into magnetite $Fe_3O_4$. No evidence is found spectroscopically of hemite $\alpha$-$Fe_2O_3$, maghemite $\gamma$-$Fe_2O_3$, or of $Bi_2O_3$. This relates to the controversy regarding the magnitude of magnetization in $BiFeO_3$.

Bismuth ferrite ($BiFeO_3$) is of considerable research interest as a lead-free ferroelectric capacitor for commercial Random Access Memories (FRAMs) [1, 2], and has been used most recently by Fujitsu for that purpose [3]. In addition, it has attracted considerable interest as a multiferroic, magnetoelectric material [4, 5]. In the latter context, there has been some controversy regarding the magnitude of its magnetization at ambient temperatures, with Wang *et al.* reporting large values of order 1 Bohr magneton per primitive cell [6], and Eerenstein *et al.* finding nearly zero [7]. A study by Bea and Bibes *et al.* revealed small amounts of $\gamma$-$Fe_2O_3$ present, and showed that this could account for the large extrinsic values in some samples [8], but the problem remains controversial. Other possible explanations exist, such as the model of Privratska and Janovec [9, 10], which shows that domain walls can be ferromagnetic in antiferromagnetic materials; we note that this implies a greater magnetization $M$ in very thin films than in thick films, due to the fact that the volume fraction of walls becomes significant in the thinner films [11]. This is in accord with experiments [11], which show an order of magnitude enhancement in *M* in the thinnest films.

---

[*] Author to whom correspondence should be addressed; email: XL217@cam.ac.uk
[§] Email: jsco99@esc.cam.ac.uk

In the present work we have used micro-Raman techniques described previously [12] to examine the phase separation of bismuth ferrite under d.c. voltage stressing at room temperature, in order to see if magnetic phases are separated in the process. As described below, we have found that magnetite is formed under these conditions. Since the Raman spectra of the $BiFeO_3$ starting material agrees very well with those recently published on virgin thin-film samples [13-17], these studies show that the magnetite is formed entirely by the applied d.c. voltage and does not exist in significant concentrations in the unstressed specimens. We note that although $Fe_3O_4$ and $Fe_2O_3$ differ by only 10% oxygen concentration, their bondings differ and this gives rise to distinguishable Raman spectra.

The bismuth ferrite film with thickness of 185 nm was made by using sol-gel spin coating techniques on a $Pt/TiO_2/SiO_2/Si$ commercialized substrate. The top electrodes (Pt dots of 100 μm x 100 μm) were then deposited by sputtering on the films via TEM grids. Micro-Raman spectra were recorded in backscattering geometry at ambient conditions by CCD collection. Two types of Raman instruments were used in the present work: RM1000 Renishaw systems which uses a 514.5 nm laser and LabRam 300 equipped with 632.8 nm laser. The electrical measurements were conducted using a Radiant RT 6000S ferroelectric system. The crystallization property of our film has been confirmed by XRD measurements.

The film was stressed by applying a d.c. voltage with increasing magnitudes. At low voltages, the film shows high resistivity. The d.c. breakdown voltage is about 13 V (i.e. $E_B$ breakdown field ~700 kV/cm,) for 2 second endurance and is relatively repeatable. The a.c. breakdown voltage is relatively higher as expected and is about 15 V (i.e. ~800 kV/cm) evaluated by hysteresis measurements on a number of electrodes at 1 kHz frequency. Pre-breakdown could be achieved after applying a d.c. voltage slightly lower than the breakdown strength. This manifests itself by an abrupt increase of leakage current and the appearance of a few degraded spots on the electrode (Fig 1, the dark spots as indicated by the arrow are used for Raman measurements shown later).

Then we carried out Raman studies on this film as described in detail previously [12]. The Raman spectra collected from a normal untreated region and a high-dc-bias-induced degraded region in a $BiFeO_3$ thin film are shown in Fig 2. One can see that the Raman spectrum collected from the untreated region shows features consistent with virgin $BiFeO_3$ thin films [13-17], e.g. peaks at 141 $cm^{-1}$, 170 $cm^{-1}$, 215 $cm^{-1}$, 261 $cm^{-1}$, 276 $cm^{-1}$, 337 $cm^{-1}$, 368 $cm^{-1}$, 467 $cm^{-1}$, 518 $cm^{-1}$ and 604 $cm^{-1}$. It has been noted [17] that the Raman spectra of $BiFeO_3$ thin films are very different from those of bulk $BiFeO_3$ [18] due to their defferent sysmetries: the structure for thin films is generally either tetragonal or monclinic, while it is rhombohedral for bulk samples [17].

Remarkably, the Raman spectra collected from the degraded areas are identical and are comparable with those for $Fe_3O_4$ phase (called magnetite) [19-23], evidenced by the strong peak at ~642 $cm^{-1}$ and several other characteristic peaks at ~314 $cm^{-1}$, 395 $cm^{-1}$, and 517 $cm^{-1}$, although these peaks are slightly shifted to lower wavenumbers compared with bulk $Fe_3O_4$. This may be due to the effect of strain on these $Fe_3O_4$ nano-particles.

Neither the ambient stable phase $\alpha$-$Fe_2O_3$ (called hematite) [21, 24], nor $\gamma$-$Fe_2O_3$ (called maghemite) [21, 25-27] is present in the degraded regions. Interestingly, $Bi_2O_3$ [28-33] seems not to be an end product found in the breakdown dendritic path in all these Bi-containing materials, such as thin-film $BiFeO_3$ (this work), $Bi_{4-x}Sm_xTi_3O_{12}$ [12], and $SrBi_2Ta_2O_9$ [34].

Note that $\gamma$-$Fe_2O_3$ has an inverse spinel structure with a tetragonal sublattice distortion. It can be seen as an iron deficient form of $Fe_3O_4$, with structural formula $Fe^{3+}_{21.33}\square_{2.67}O^{2-}_{32}$, where $\square_{2.67}$ means 2.67 vacancies in octahedral site of the spinel structure [35].

Both $Fe_3O_4$ and $\gamma$-$Fe_2O_3$ are well known ferrimagnetic materials. The Curie temperature is about 580 °C for magnetite ($Fe_3O_4$), and about 600 °C for maghemite ($\gamma$-$Fe_2O_3$). In particular, magnetite ($Fe_3O_4$) accounts for the most magnetic of all the naturally occurring minerals on the Earth. So we suggest that these results may relate to the controversy regarding the magnitude of magnetization in $BiFeO_3$ thin films. This point becomes a particular concern, considering the high coercive field $E_c$ observed for $BiFeO_3$ thin films: it has been shown that the coercive field in bismuth ferrite thin films is about 300 kV/cm, twice or three times larger than that in common ferroelectric thin films (<100 kV/cm) [13]; hysteresis measurements sometimes require an electric field as large as 1~2 MV/cm [13, 36-38], compatible or even larger than the field we applied in this work. Under such a high field, interfacial phase separation of $BiFeO_3$ into magnetite is reasonably expected due to intensive charge injection from the electrode.

Note that $BiFeO_3$ single crystal or ceramics show space-commensurate ferroelectricity below the Curie point (1103 K) and space-incommensurate antiferromagnetism below the Neel temperature (643 K), with a superimposed cycloidal modulation with a wavelength of ~62 nm [39]. Consequently, its remanent magnetization and potential magnetoelectric effect both vanish macroscopically, leading to a weak quadratic magnetoelectric behavior rather than a strong linear magnetoelectric behavior [4, 5]. It has been found that the cycloidal order can be destroyed and the non-zero remanent magnetization (and potential magnetoelectric effect) can be released by applying a sufficiently large magnetic field of 20 T in bulk $BiFeO_3$ [40] or by chemical substitutions in thin films [41]. In this work, we show that phase separation of $BiFeO_3$ into nano-regions of magnetite under a d.c. field, comparable to the applied field for switching measurement, may also lead to a possibility that electrical testing of samples might itself produce extrinsic magnetization.

The presence of decomposition at room temperature at voltages and fields near that of the coercive switching voltages and fields has implications for the Fujitsu BiFeO$_3$:Mn FRAM memories.

In summary, we show that under ~700 kV/cm of dc voltage stressing for only a few seconds, bismuth ferrite BiFeO$_3$ thin film phase separates into magnetite. However, there is no evidence found spectroscopically of hemite (α-Fe$_2$O$_3$), maghemite (γ-Fe$_2$O$_3$) or any polymorphs of Bi$_2$O$_3$. Our work may relate to the controversy regarding the magnitude of magnetization in BiFeO$_3$.

X.J.L would like to thank Cambridge Overseas Trust for supporting during this work. We are grateful to Dr. Anna Moisala and Dr. Zoe Barber for help with the Raman measurements.

Figure captions :

Fig 1 (color online) Optical image of the electrode after applying a high d.c. voltage.

Fig 2 (color online) micro-Raman spectra recorded from the normal regions and the dendritic degraded regions formed by applying a large dc bias in a BiFeO$_3$ film. The inset presents a magnified view of the spectra from ~250 cm$^{-1}$ to 800 cm$^{-1}$.

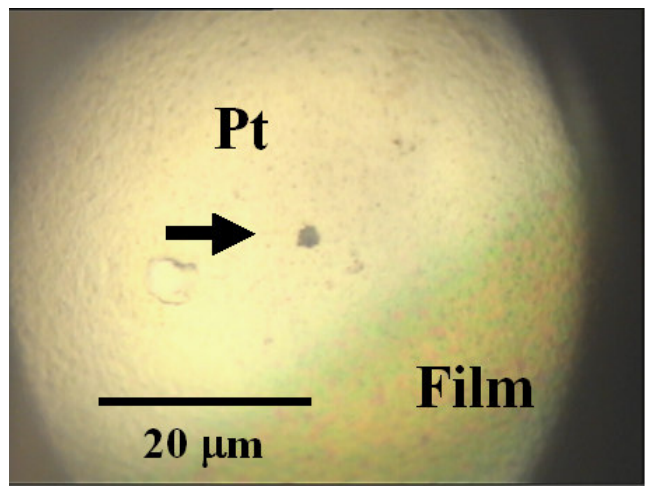

Fig 1

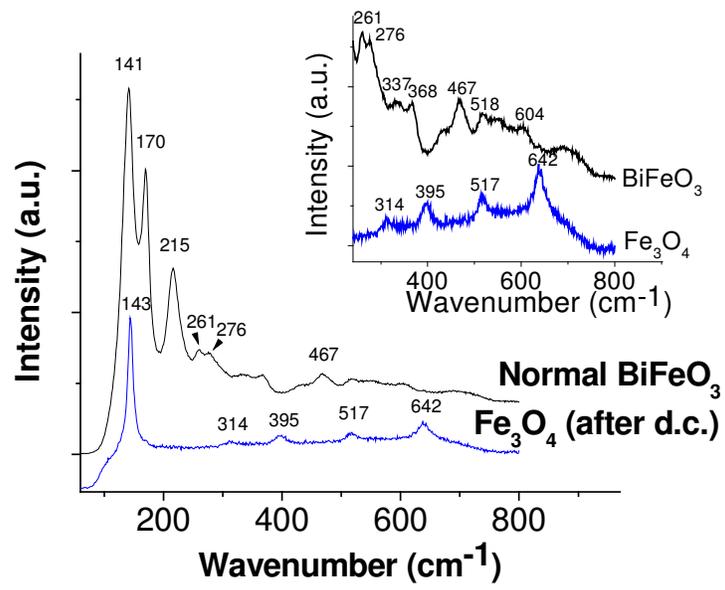

Fig 2